# Magnetometer Based On Spin Wave Interferometer

M. Balynsky[1], D. Gutierrez[1], H. Chiang [1], A. Kozhevnikov[2,3], Y. Filimonov[2,3], A.A. Balandin[1], and A. Khitun[1]

[1]*Department of Electrical and Computer Engineering, University of California - Riverside, Riverside, California, USA 92521*

[2]*Kotelnikov Institute of Radioengineering and Electronics of the Russian Academy of Sciences, Saratov, Russia 410019*

[3]*Saratov State University, Saratov, Russia 410012*

**Abstract:** We describe magnetic field sensor based on spin wave interferometer. Its sensing element consists of a magnetic cross junction with four micro-antennas fabricated at the edges. Two of these antennas are used for spin wave excitation and two others antennas are used for the detection of the inductive voltage produced by the interfering spin waves. Two waves propagating in the orthogonal arms of the cross may accumulate significantly different phase shifts depending on the magnitude and the direction of the external magnetic field. This phenomenon is utilized for magnetic field sensing. The sensitivity has maximum at the destructive interference condition, where a small change of the external magnetic field results in a drastic increase of the inductive voltage as well as the change of the output phase. We report experimental data obtained on a micrometer scale $Y_3Fe_2(FeO_4)_3$ cross structure. The change of the inductive voltage near the destructive interference point exceeds 40 dB per 1 Oe. At the same time, the phase of the output exhibit a π-phase shift within 1 Oe. The data are collected for three different orientations of the sensor in magnetic field at room temperature. Taking into account low thermal noise in ferrite structures, the maximum sensitivity of spin wave magnetometer may exceed atta Tesla. Other appealing advantages include compactness, fast data acquisition and wide temperature operating range. The physical limits of spin wave interferometers are also discussed.

# I. Introduction.

Magnetometers are among the most widely used instruments in a variety of applications [1]. There are different types of magnetic sensors including Superconducting Quantum Interference Devices (SQUID) [2], resonance magnetometers (e.g. Proton magnetometer) [3], $He^4$ $e^-$-spin magnetometer [4], solid state magnetometers (e.g. Fluxgate, Giant Magneto-Impedance, Magneto-Resistive, Hall, Magneto-Electric [5]),and the variety of fiber optic magnetometers [6-8]. The operation of the above-mentioned magnetometers is based on different physical processes offering unique advantages for the each type of magnetic sensor. Sensitivity, intrinsic noise, volume, energy budget, and cost are the most important magnetometer characteristics.

The sensitivity of the magnetic sensor also known as the transfer function is the characteristic which relates the input magnetic field to the output voltage $S_B^V(V/T)$ [9]. The most sensitive sensors show the transfer coefficient as high as $10^5$ V/T [9]. The intrinsic noise of the sensor $B_n(f)$ is the second important parameter, where $f$ is the frequency. The intrinsic noise is usually estimated by measuring the time variation of the output voltage of the sensor followed by the Fourier transform. Then, the result is divided by the transfer function $S_B^V(V/T)$ leading to $B_{n,eq}(f)$ expressed in $T/\sqrt{Hz}$ [9]. As for today, SQUID magnetometers demonstrate the highest sensitivity enabling the detection of extremely subtle magnetic fields as low as 5 aT ($5×10^{−18}$ T) with noise level of $3x10^{−15}\, T/\sqrt{Hz}$ [10]. However, the maximum sensitivity of SQUIDs is achieved at the cryogenic temperatures, which translates in a high cost and narrows SQUIDs practical applications. In contrast, solid state magnetometers are much less expensive, compact and can operate at room temperature. The latter stimulates the search for highly sensitive and room temperature operating solid state sensors.

One of the promising route toward highly sensitive solid state magnetometers was proposed in Ref. [11]. It was demonstrated the prototype for measuring low alternating magnetic fields by means of ferrite-garnet films with planar anisotropy. The initial experiments were carried out with Bi-containing RE ferrite-garnet $(BiLuPr)_3(FeGa)_5O_{12}$ films enabling detection of $10^{-7}$ Oe magnetic field. Later on, the same group of authors demonstrated a prototype based on epitaxialy grown yttrium iron garnet (YIG) films [12]. The prototype of the 3D YIG magnetometer was experimentally tested, demonstrating the detection level below as $10^{−12}\, T/\sqrt{Hz}$ at frequencies above 0.1 Hz. The minimum noise level was projected at the level $10^{−15}\, T/\sqrt{Hz}$ at room temperature [12]. The high sensitivity of the YIG-based sensor is mainly due to the low intrinsic noise, which makes this material a perfect candidate for magnetic field sensing.
.
In this work, we present magnetic field sensor based on spin wave interferometer. The rest of the paper is organized as follows. In Section II, we describe the material structure and the principle of operation of the sensor. In Section III, we present experimental data obtained for a micrometer scale prototype based on YIG structure. The discussion and conclusions are given in Sections IV and V, respectively.

.

## II. Material Structure and Principle of Operation

The schematic of the sensor are shown in Figure 1. Sensing element is a magnetic cross made of a material with low spin wave damping (e.g. yttrium iron garnet $Y_3Fe_2(FeO_4)_3$). It is a four terminal device, where the terminals are micro-antennas fabricated on the edges of the cross (e.g. Π-shaped antennas). The antennas are directly placed on the top of the cross. Two of these antennas (i.e. marked as 1 and 2) are used for spin wave excitation, and the other two (i.e. marked as 3 and 4) are used for the spin wave detection via the inductive voltage measurements [13]. Spin wave generating antennas are connected to the same RF source via the set of phase shifters and attenuators. The output antennas are connected to the detector. The cross structure is placed on top of the magnetic substrate, which is aimed to provide a DC bias magnetic field (e.g. in-plane magnetic field directed along a virtual line connecting antennas 1 and 3).

The principle of operation is the following. Input spin waves are excited by passing a RF current through the antennas 1 and 2. AC electric current generates an alternating magnetic field around the current carrying wires and excite spin waves in the magnetic material beyond the antennas. The details of spin wave excitation by micro-antennas can be found elsewhere [14, 15]. Spin wave propagate through the cross structure and reach the output ports. The propagating waves alter the magnetic flux from the structure and induce an inductive voltage $V_{ind}$ in the output antenna. The output voltage has maximum when spin wave are coming in phase (i.e., constructive interference). The output voltage has minimum when the waves are coming out-of-phase (i.e., destructive interference) as illustrated in the inset to Fig.1. The phase difference among the waves depends on the external magnetic field $H$, which may produce significantly different phase shifts for the spin waves propagating in the orthogonal arms. Thus, the output voltage depends on the external magnetic field. As we will show later, the maximum sensitivity to the magnetic field occurs in the case of destructive interference. In this case, a small phase difference produced by the external field variation results in significant increase of the output inductive voltage. Below, we describe the physical model of the sensor and estimate its transverse characteristic $\partial V/\partial H$.

Spin wave is a propagating disturbance of magnetization in ordered magnetic materials [16], which can be described as a sum of the $\vec{M}_0$ static and the dynamic $\vec{m}(r,t)$ components ($|m| \ll |M|$) as follows [17]:

$$\vec{M}(r,t) = \vec{M}_0 + \vec{m}(r,t),$$

$$m(r,t) = m_0 \cdot exp[-\kappa r] \cdot sin(k_0 r - \omega t + \varphi_0 ), \tag{1}$$

where the dynamic part is as a propagating wave, $m_0$ and $\varphi_0$ are the initial amplitude and phase, $\kappa$ is the damping constant, $r$ is the distance traveled, $k_0$ is the wave vector, $\omega$ is the frequency $\omega = 2\pi f$, and $t$ is the time. Spin wave excited at ports 1 and 2 have the same frequency $\omega$, which is defined by the frequency of the input RF signal. The initial amplitudes and phases of the generated spin waves wave are controlled by the system of phase shifters and attenuators. Generated spin waves propagate through the cross junction and reach the output antennas (e.g. antenna 3). The disturbance of magnetization at the output is a result of the spin wave interference

$$\vec{m}(l,t) = \vec{m}_1(l,t) + \vec{m}_2(l,t), \tag{2}$$

where $\vec{m}_1(l,t)$ and $\vec{m}_2(l,t)$ are the dynamic components of spin waves generated at ports 1 and 2, respectively; $l$ is the distance traveled. The set of attenuators is needed to equalize the amplitudes of the two spin waves coming to the output $|\vec{m}_1(l,t)| = |\vec{m}_2(l,t)|$. In this case, the amplitude of the magnetization change caused by the spin wave interference can be found as:

$$m(l,t) = m_0 \cdot exp[-\kappa l] \cdot \sqrt{2 + 2\cos\Delta\varphi} \cdot \sin(\omega t + \theta), \tag{3}$$

where $\Delta\varphi$ is the phase difference between the interfering waves, $\theta = \text{atan}(\sin\Delta\varphi/(1+\cos\Delta\varphi)$. The phase difference $\Delta\varphi$ is a sum of two parts:

$$\Delta\varphi = \Delta\varphi_0 + \Delta\varphi(H), \tag{4}$$

where $\Delta\varphi_0$ is the difference in the initial phases, and $\Delta\varphi(H) = \varphi_1(H) - \varphi_2(H)$ is the phase difference which arises during the spin wave propagation. The phase shift accumulated by *i-th* (*i*=1,2) spin wave during propagation is given by

$$\varphi_i(H) = \int_0^l k_i(\vec{H})dr, \tag{5}$$

where the particular form of the wavenumber $k(\vec{H})$ dependence varies for magnetic materials, film dimensions, the mutual direction of wave propagation and the external magnetic field [18]. For example, spin waves propagating perpendicular to the external magnetic field (magnetostatic surface spin wave – MSSW) and spin waves propagating parallel to the direction of the external field (backward volume magnetostatic spin wave – BVMSW) may obtain significantly different phase shifts for the same field. The phase shift $\Delta\phi$ produced by the external magnetic field variation $\delta H$ in the ferromagnetic film can be expressed as follows[19]:

$$\frac{\Delta\varphi}{\partial H} = \frac{l}{d}\frac{(\gamma H)^2 + \omega^2}{2\pi\gamma^2 M_S H^2}, \quad k \parallel H \qquad \text{(BVMSW)}$$

$$\frac{\Delta\varphi}{\partial H} = -\frac{l}{d}\frac{\gamma^2(H+2\pi M_S)}{\omega^2-\gamma^2 H(H+2\pi M_S)}, \quad k \perp H \text{ (MSSW)} \tag{6}$$

where $\Delta\varphi$ is the phase shift produced by the change of the external magnetic field $\delta H$, $d$ is the thickness of the waveguide, $\gamma$ is the gyromagnetic ratio, $4\pi M_s$ is the saturation magnetization of the magnetic material. The formula are derived for the approximate dispersion law and valid for $\delta H \ll H$ [20].

Propagating spin waves alters the magnetic flux $\Phi_m$ from the structure, which results in the inductive voltage $V_{ind}$ according to the Faraday's law of induction:

$$V_{ind}(t) = -\frac{d\Phi_m}{dt} = \omega \Upsilon \frac{dm(l,t)}{dt}, \tag{7}$$

where $\Upsilon$ is a constant parameter, which accounts for the geometry and material properties of the antenna (e.g. the area and the shape of the antenna contour, antenna's resistance) [13]. The average output voltage can be found as:

$$\bar{V}_{ind} = \bar{V}_0 \cdot \sqrt{2 + 2\cos\Delta\varphi}, \tag{8}$$

where $\bar{V}_0$ is the average inductive voltage produced by just one spin wave generating antenna. The set of Eqs.(1-8) connect the output inductive voltage to the phase difference among the interfering spin waves, which, in turn, relates it to the external magnetic field.

In order to find the regions of parameters providing maximum sensitivity, we present the results of numerical modeling. The response characteristic of the proposed sensor $\partial V/\partial H$ is defined by the two major factors: (i) spin wave phase sensitivity to the external magnetic field, and (ii) minimum phase shift which can be detected via the inductive voltage measurements:

$$\frac{\partial V}{\partial H} = \frac{\partial V}{\partial(\Delta\varphi)} \cdot \frac{\partial(\Delta\varphi)}{\partial H}. \tag{9}$$

In Fig. 2(A), we present the results of numerical simulations showing the change of the output inductive voltage as a function of the phase difference between the two interfering spin waves $\partial V/\partial(\Delta\varphi)$. According to Eq. (8), the maximum change of the inductive voltage occurs in the case of the destructive wave interference $\Delta\varphi = \pi$, where a small change of the phase difference results in a drastic increase of the inductive voltage.

In Fig.2(B), we present the results of numerical modeling showing the phase change accumulated by the propagating spin wave due to the magnetic field variation $\partial\varphi/\partial H$ according to Eq.(6). The material parameters used in numerical simulations are the following: 4πM$_s$ = 1750 Gs; γ = 2π·2.82 MHz/Oe; $l/d$ = 960; $f$ = 4.95 GHz. The black and the red curves in Fig. 2(B) show the phase sensitivity of the BVMSW and MSSW type of waves respectively. According to Eqs.(6), the maximum of BVSM sensitivity corresponds to H=0, while the maximum sensitivity of the MSSW waves corresponds to the case $\omega^2 = \gamma^2 H(H + 2\pi M_S)$. It is important to note that the maximum sensitivity can be achieved only for one type of propagating waves at a time. More than that, there is only a finite frequency overlap where the both types of spin waves can propagate. The overlap occurs due to the effect of the shape anisotropy in the cross junction [21]. At the chosen material parameters and operational frequency $f$ = 4.95 GHz, the overlap takes place around H=1100 Oe as shown in Fig.2(B). The width of the overlap is about 110 Oe. The inset in Fig.2(B) shows the phase sensitivity $\partial\varphi/\partial H$ of MSSW and BVMSW type of waves: $\partial\varphi/\partial H > 0$ for MSSW, and $\partial\varphi/\partial H < 0$ for BVMSW. This asymmetry in the phase change is important to magnetometer functionality. As we show in the next Section, the proposed magnetometer makes it possible to detect not only the change in the magnitude of magnetic field but also its direction.

**III. Experimental Data**

The photo of the sensing element and connection schematics are shown in Fig. 3. The element is a cross junction made of single crystal $Y_3Fe_2(FeO_4)_3$ film. The film was grown on top of a Gadolinium Gallium Garnett ($Gd_3Ga_5O_{12}$) substrate using the liquid-phase epitaxy technique. The micro-patterning was performed by laser ablation using a pulsed infrared laser (λ≈1.03 μm), with a pulse duration of ~256 ns. The YIG cross has the following dimension: the length of the each waveguide is 3.65 mm; the width is 650 μm; and the YIG film thickness is 3.8 μm; and saturation magnetization of $4\pi M_0 \approx 1750\ Oe$. There are four Π-shaped micro-antennas fabricated on the edges of the cross. Antennas were fabricated from a gold wire of thickness 24.5μm and placed directly at the top of the YIG surface. The antennas are connected to a programmable network analyzer (PNA) Keysight N5241A. Two of the antennas marked as 1 and 2 in Fig.3 are used to generate two input spin waves. The inductive voltage is detected by the antennas marked as 3 and 4. The details of the inductive measurement technique can be found elsewhere [22]. There is a set of attenuators (PE7087) and a phase shifters (ARRA 9428A) to independently control input power and the phase of the spin wave signals generated at the input ports 1 and 2. The device was placed inside an electromagnet to control the bias magnetic field from −2000 Oe to +2000 Oe. In general, the sensor does not require a controllable source of magnetic field as it operates at some constant magnetic field (e.g. bias magnetic field provided by the substrate as shown in Fig.1). The electromagnet is needed for the test experiments aimed to identify the most robust regions of operation. Before the experiment, we determined the region in the frequency-bias magnetic field space where both types of waves BVMSW and MSSW can propagate as described in [21]. The latter is critically important for the operation of cross-shape devices because the input spin waves initially propagate perpendicular to the bias field (MSSW), whereas they reach the output by propagating along the external magnetic field (BVMSW). The most prominent overlap takes place in the frequency range from 4 GHz to 5 GHz and bias magnetic field from 750 Oe to 1200 Oe.

The experimental procedure includes two major steps. First, we use the system of attenuators and phase shifters to ensure the destructive spin wave interference at one of the output ports (e.g., antennas 3 or 4). The amplitudes of the spin waves coming to the output port are equalized by the attenuators. Then, we measure the output voltage for the phase difference between the interfering spin waves from $0\pi$ to $2\pi$, where the phase difference is controlled by the phase shifters. The minimum of the inductive voltage corresponds to the destructive spin wave interference. Second, we vary the strength of the bias magnetic field ±1 Oe and measure the change of the output inductive voltage in the vicinity of the destructive interference point.

We carried out three sets of experiments aimed to show the change of the output inductive voltage with respect to the changing magnetic field at different directions of the magnetic field. The operational frequency *f* is 4.95 GHz, and the bias magnetic field is *H* = 1074 Oe in all cases. The input microwave power at ports 1 and 2 is -6 dBm (0.5 mW). All experiments are done at room temperature. In Fig. 4, we present experimental data for the bias magnetic field *H* directed parallel to the virtual line connecting ports 1 and 3 as illustrated in the inset. Fig.4(A) shows the amplitude (red markers) and the phase (blue markers) of the inductive voltage detected at port 3. On can clearly see the result of spin wave interference, which provides maximum output voltage about 9 mV in case of constructive spin wave interference (i.e. phase difference between the interfering spin waves $\Delta\varphi_{12} = 0\pi, 2\pi$). The output has minimum about 10 μV in case of the destructive spin wave interference (i.e., $\Delta\varphi_{12} = 1\pi$). This is the most sensitive regime of operation according to the physical model described in the previous Section II. We fix the position of phase shifters and attenuators to keep the sensing element in the destructive interference regime. Next, we vary the strength of the bias magnetic field. In Fig.4(B), we present experimental data showing the output voltage detected caused by the magnetic field variation in 1 Oe. The data show a prominent signal change about 40 dB per 1 Oe.

Similar experiments were accomplished for two different orientations of the bias magnetic field. Fig.5(A) shows the amplitude (red markers) and the phase (blue markers) of the inductive voltage detected at port 3. The bias magnetic field is directed perpendicular to the virtual line connecting ports 1 and 3. The output voltage has maximum about 4.2 mV corresponding to the constructive spin wave interference. The minimum voltage about 1 µV is detected in the case of destructive spin wave interference. The change of the output voltage due to the variation of the bias magnetic field is shown in Fig.5(B). The amplitude of the output voltage increases drastically (i.e., more than 50 dB) with the 1 Oe change of the bias magnetic field. Finally, we repeated measurements for bias magnetic field directed at $45^0$ with respect to the virtual line connecting ports 1 and 3. Experimental data are shown in Fig.6(A) and Fig.6(B), respectively. There are two important observations we want to outline for the $45^0$ case: (i) the inductive voltages are approximately the same ($\pm$ 1dB) at ports 3 and 4; (ii) the relative change of the output voltage within the destructive interference point is relatively small (i.e., 5 dB per 1 Oe) compare to the previous two experiment configurations. The accuracy of the inductive voltage measurements is $\pm$ 0.00046 mV.

## IV. Discussion

Experimental data presented above demonstrate prominent amplitude as well as the phase change of the output inductive voltage in the vicinity of the destructive spin wave interference. In this Section, we estimate sensor sensitivity and discuss its potential advantages and limits. The transfer function $S_B^V$ (V/T) is the key sensor characteristic which relates the input (magnetic field) to the output (inductive voltage). The maximum value about 1 mV per 1 Oe was detected for in-plane magnetic field directed along the virtual line connecting ports 1 and 3 as shown in Fig.5. Note that the experimental data were obtained on the first, non-optimized prototype. For instance, the input power for the spin wave generating antennas was only -6 dBm (0.5 mW). The transfer function can be further enhanced by applying a higher input power. The same YIG device can sustain operation at higher input power level of 0 dBm (10 mW). There are certain limits on pumping energy into spin wave signals, which results in dynamic instabilities [23]. In Fig.7, we present experimental data showing the dependence of the transmitted signal on the input power. The instability restricts the input power of the YIG prototype at the level about +1 dBm. However, the relative change of the output inductive voltage exceeds 40 dB/ 1 Oe. Using standard low noise electronics (e.g. an operational amplifier), it may be possible to amplify the output inductive voltage over the orders of magnitude [24].

The second important specification is the intrinsic noise of the sensor, conveniently referred at the sensor's input, in units of the square root of an equivalent magnetic power spectrum, $B_{n,eq}(f)$ [9]. Usually, the time variations of the output voltage are recorded, followed by a Fourier transform, and then divided by the transfer function $S_B^V$ to get $B_{n,eq}(f)$, expressed in T/√Hz [9]. According to the formalism described [12], the spectral density of an effective external magnetic field that describes thermal fluctuations in a medium with magnetic losses can be evaluated as the intrinsic thermal noise of magnetic material can be estimated as follows:

$$B_n(f) = \sqrt{\frac{T\Delta H}{4\pi V\gamma M_S^2}} \quad , \tag{10}$$

where $T$ is an absolute temperature in erg and $V$ is the magnetic film volume. The estimates for a YIG sample with the following parameters $\Delta H$ = 1 Oe, $M$ = 140 Oe, $V$ = 2.5 × 10⁻⁴ cm³, frequency band 1 Hz, show $10^{-15}\, T/\sqrt{Hz}$ at room temperature [12]. Also, we want to note that the output of the spin wave-based magnetometer is AC inductive voltage, which allows us to exploit a phase lock-in amplifier [25]. Taking into account the RF operating range and well-defined output frequency, signals up to 1 million times smaller than noise level can be detected. Altogether, it makes feasible to reach the detection level of cooled SQUIDs (i.e., atta Tesla) but with YIG sensors operating at room temperature.

Detecting the output phase provides us an alternative way for magnetic field sensing. The phase of the output exhibits a180⁰ abrupt jump near the destructive interference point. The jump occurs with less than 1 Oe variation of the bias magnetic field. Thus, the change of the magnetic field is related to the change of the output phase. In this scenario, the maximum field sensitivity is defined by the precision of the phase measurement. A sub-micro-degree phase measurement technique was demonstrated for lock-in amplifiers [26]. In this case, the ultimate sensitivity of the interferometer combined with a phase lock-in amplifier may exceed 10⁻¹³ Tesla. There are several advantages of using phase measurement compare to the amplitude-based approach. The phase change does not depend on the amplitudes of the input signals, which allows us to minimize power consumption. The sign of the phase change is directly related to the decrease/increase of the magnetic field while the amplitude of the output is almost symmetric (e.g., as shown in Figs. 4-6). The combination of the amplitude and phase measurements makes it possible to detect the change in the amplitude and the direction of the sensing magnetic field.

Fast data accusation is another appealing property of the proposed magnetometer. The time delay is limited by two factors: spin wave propagation time, and output voltage averaging time. Spin wave propagation time in the millimeter-long prototype is about 0.1 µs. Then, it will take at least one period of oscillation (0.1 ns) to detect the average amplitude of the output signal. Scaling down the size of the sensing element is in favor of the proposed magnetometer by reducing its time delay, increasing operation frequency, and minimizing thermal noise. Potentially, it is possible to build interferometers operating with exchange spin waves with the length of the sensing element in a deep sub-micrometer range. For instance, the time delay in a 500 nm structure with a velocity of exchange magnons of $10^3$ m/s is only 0.5 ns [27]. It is also important to note that miniaturization reduces propagation losses [27]. At the same time, the frequency of exchange magnons achieves 7 THz at the edge of the first Brillouin zone [28]. The latter translates in an intrigue possibility of building highly sensitive magnetic magnetometers capable of detecting $10^{-18}$ T change in the fast varying magnetic field with maximum frequency up to 1 GHz.

Sensing elements made of YIG or similar ferrite materials possess a wide temperature operating range from cryogenic to the 560K. One more appealing property of the proposed magnetometer is that a number of interferometers can be combined in a one detector network. It is possible to build a network comprising a large number of interferometers in a phased array in which the relative phases (amplitudes) of input spin waves are adjusted in such a way that the incoming magnetic signals of interest are amplified while other signal coming from undesired directions are suppressed. Phased arrays are currently used in a variety of applications. For example, the MESSENGER spacecraft (mission to the planet Mercury arrived 18 March 2011) was the first deep-space mission with a phased-array antenna for communications [29]. Potentially, spin wave interferometers can be arranged in a magnetic telescope for the outer space exploration.

The need in the bias magnetic field is the main technological disadvantage of the described magnetometer. The most sensitive regime of operation requires an accurate adjustment of the bias magnetic field and the operational frequency to ensure the destructive spin wave interference at the one of the outputs. This issue can be resolved by implementing magnetic materials with out-of-plane magnetization and/or utilization of antiferromagnetic materials. The lack of experimental data on spin wave interference in such materials restricts us from a more qualitative analysis. The major magnetometer characteristics such as time delay, signal bandwidth, magnetic field sensing range are directly related to the spin wave dispersion and attenuation, which define the physical limits of using spin waves for magnetic field sensing. There is a tradeoff between material parameters (i.e. $M_s$, $\gamma$), element geometry (i.e. *l/d* ratio) and the operating range (i.e. *f*, *H*). For example, the phase sensitivity can be enhanced by increasing the *l/d* ratio. However, the increase of the propagation length *l* is associated with the signal damping and the transfer characteristics degradation.

There is a variety of ways for sensing element improvement. For example, the geometry of the cross junction can be optimized to ensure a wider overlap among the MSSW and BVMSW signals. The integration of spin wave interferometers in a network is one of the promising directions for further research. The most critical questions are related to the level of thermal noise (i.e., phase noise) in the ferrite micro and nanostructures, which remains mainly unexplored. This work is aimed to expose the idea of a magnetometer based on spin wave interference and outline its potential advantages.

**V. Conclusions**

We described magnetic field sensor based on spin wave interferometer. The principle of operation is based on the effect of magnetic field on spin wave propagation. The effect is increased by implementing spin wave interferometer, where the change of the external magnetic field affects the inductive voltage produced by the spin wave interference. We presented experimental data obtained on a micrometer scale $Y_3Fe_2(FeO_4)_3$ cross structure. The data show the change of the output inductive voltage at different orientations of the sensing element in external magnetic field. The maximum transfer characteristic exceeds 40 dB per 1 Oe at Room Temperature. Potentially, spin wave-based magnetometers may compete with SQUIDs in sensitivity due to the low thermal noise in ferrite structures. The other advantages include compactness, fast data accusation, and wide temperature operating range.

**Acknowledgement**

This was supported by the Spins and Heat in Nanoscale Electronic Systems (SHINES), an Energy Frontier Research Center funded by the U.S. Department of Energy, Office of Science, Basic Energy Sciences (BES) under Award # SC0012670.

**Figure Captions**

Figure 1. Schematics of the sensing element. It is a spin wave interferometer build on magnetic cross junction. There are four micro-antennas fabricated on the edges of the cross. Two of these antennas (i.e. ports 1 and 2) are used for spin wave excitation, and the other two (i.e. ports 3 and 4) are used for the spin wave detection via the inductive voltage measurements. Sensing element is placed on top of the magnetic substrate, which is aimed to provide a DC bias in-plane magnetic field.

Figure 2. Results of numerical simulations. (A) Output inductive voltage as a function of the phase difference between the two interfering spin waves. (B) Phase sensitivity of the propagating spin waves to the external magnetic field variation. The material parameters are the following. $4\pi M_s$=1750Gs, $\gamma$=2π·2.82MHz/Oe; $l/d$=960. The black and the red curves show the phase sensitivity of BVSM and MSSW types of waves, respectively.

Figure 3. Schematics of the experimental setup. The sensing element is a cross junction made of single crystal $Y_3Fe_2(FeO_4)_3$ film. The YIG cross has the following dimension: the length of the each waveguide is 3.65 mm; the width is 650 µm; and the YIG film thickness is 3.8 µm. There are four Π-shaped micro-antennas fabricated directly on the surface of YIG at the edges of the cross. The antennas are connected to a programmable network analyzer (PNA) Keysight N5241A. There is a set of attenuators (PE7087) and a phase shifters (ARRA 9428A) to independently control input power and the phase of the spin wave signals generated at the input ports 1 and 2. The device is placed inside an electromagnet to control the in-plane bias magnetic field from −1000 Oe to +1000 Oe.

Figure 4. Experimental data obtained for magnetic field directed along the virtual line connected ports 1 and 3. (A) The amplitude (red markers) and the phase (blue markers) of the inductive voltage detected at port 3 as a function of the spin wave phase difference. Output voltage maxima correspond to the constructive spin wave interference. The minimum of the inductive voltage corresponds to the destructive spin wave interference. (B) Output voltage dependence on the magnetic field in the vicinity of the destructive interference point.

Figure 5. Experimental data obtained for magnetic field directed along the virtual line connected ports 2 and 4. (A) The amplitude (red markers) and the phase (blue markers) of the inductive voltage detected at port 3 as a function of the spin wave phase difference. (B) Output voltage dependence on the magnetic field in the vicinity of the destructive interference point.

Figure 6. Experimental data obtained for magnetic field directed at $45^0$ to the virtual line connected ports 1 and 3. (A) The amplitude (red markers) and the phase (blue markers) of the inductive voltage detected at port 3 as a function of the spin wave phase difference. (B) Output voltage dependence on the magnetic field in the vicinity of the destructive interference point.

Figure 7. Experimental data showing the transmission characteristic as a function of input power. The transmission in YIG prototype drops due to the spin wave instability at the input power exceeding 0 dBm.

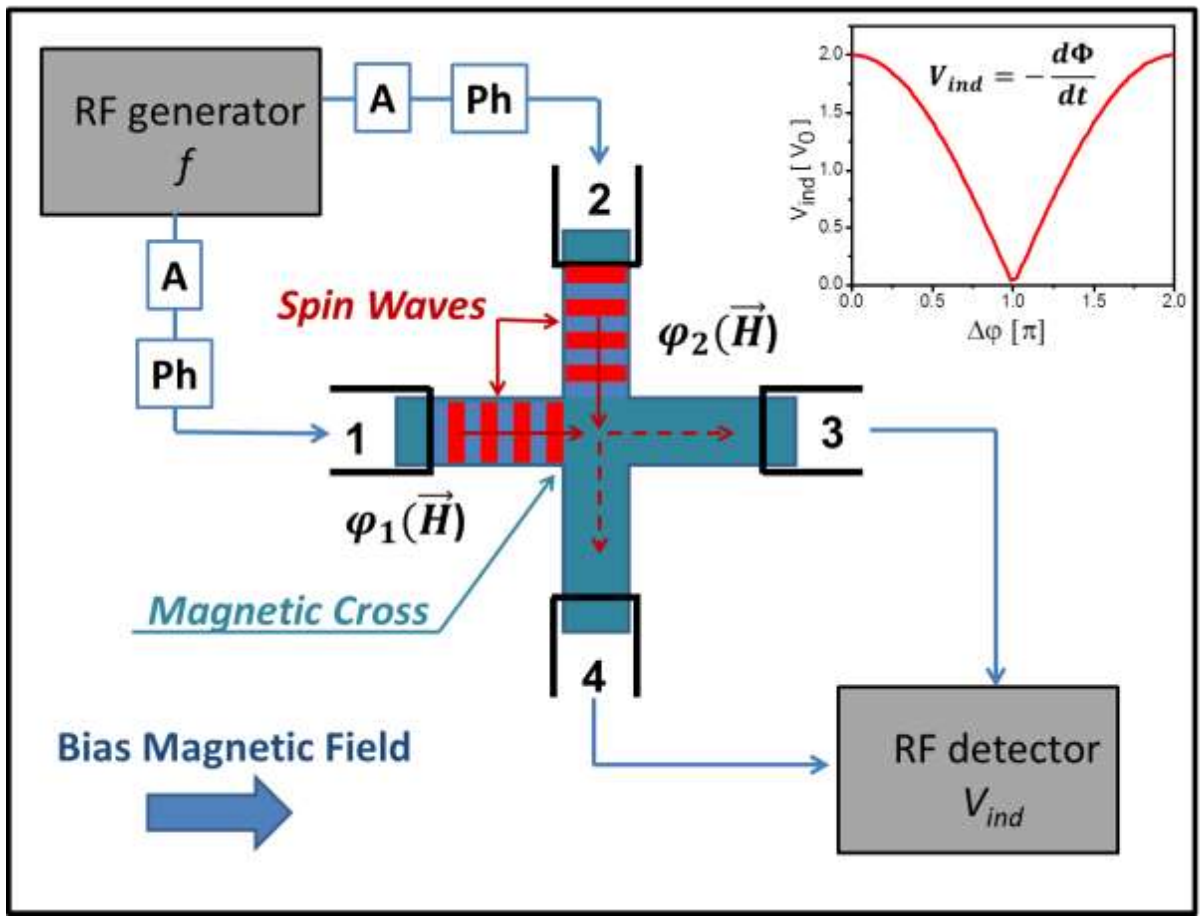

Figure 1

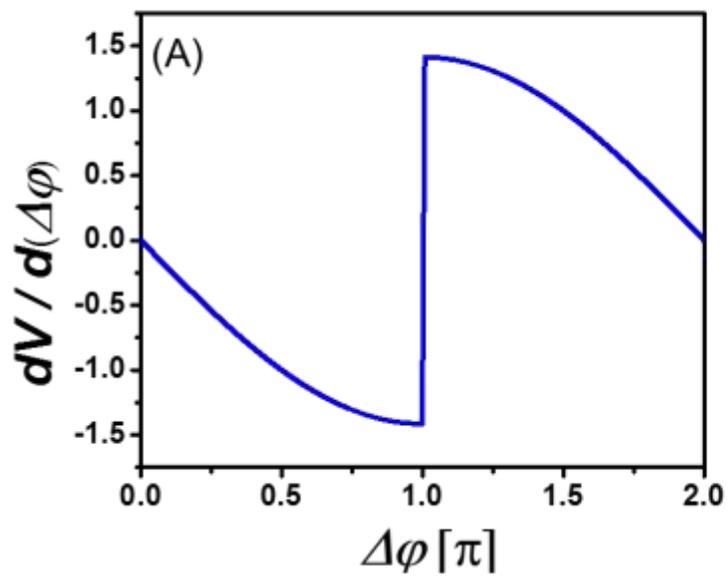 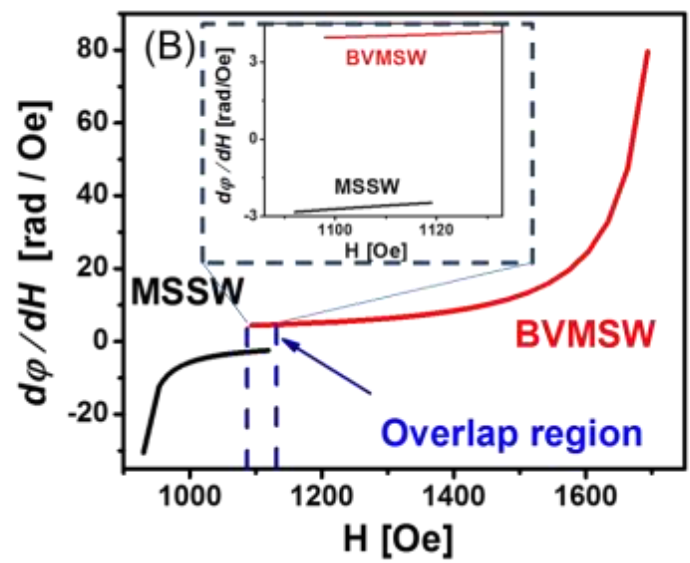

Figure 2

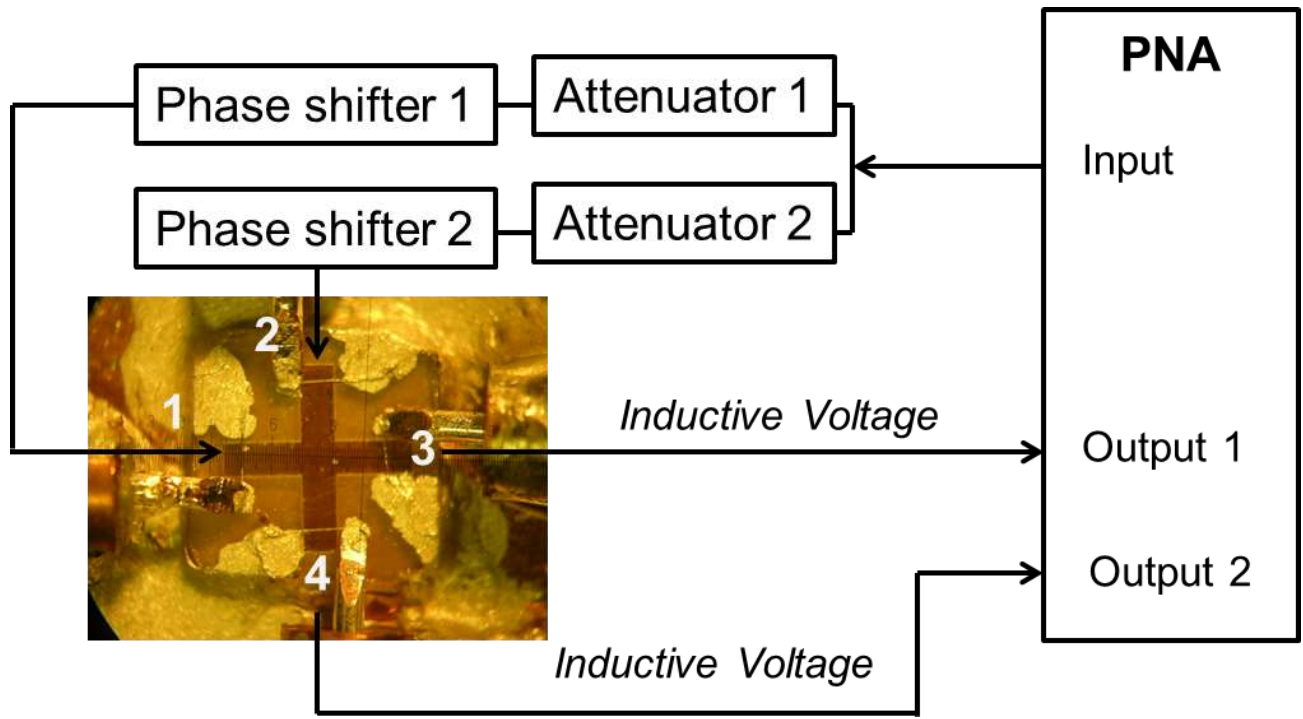

Figure 3

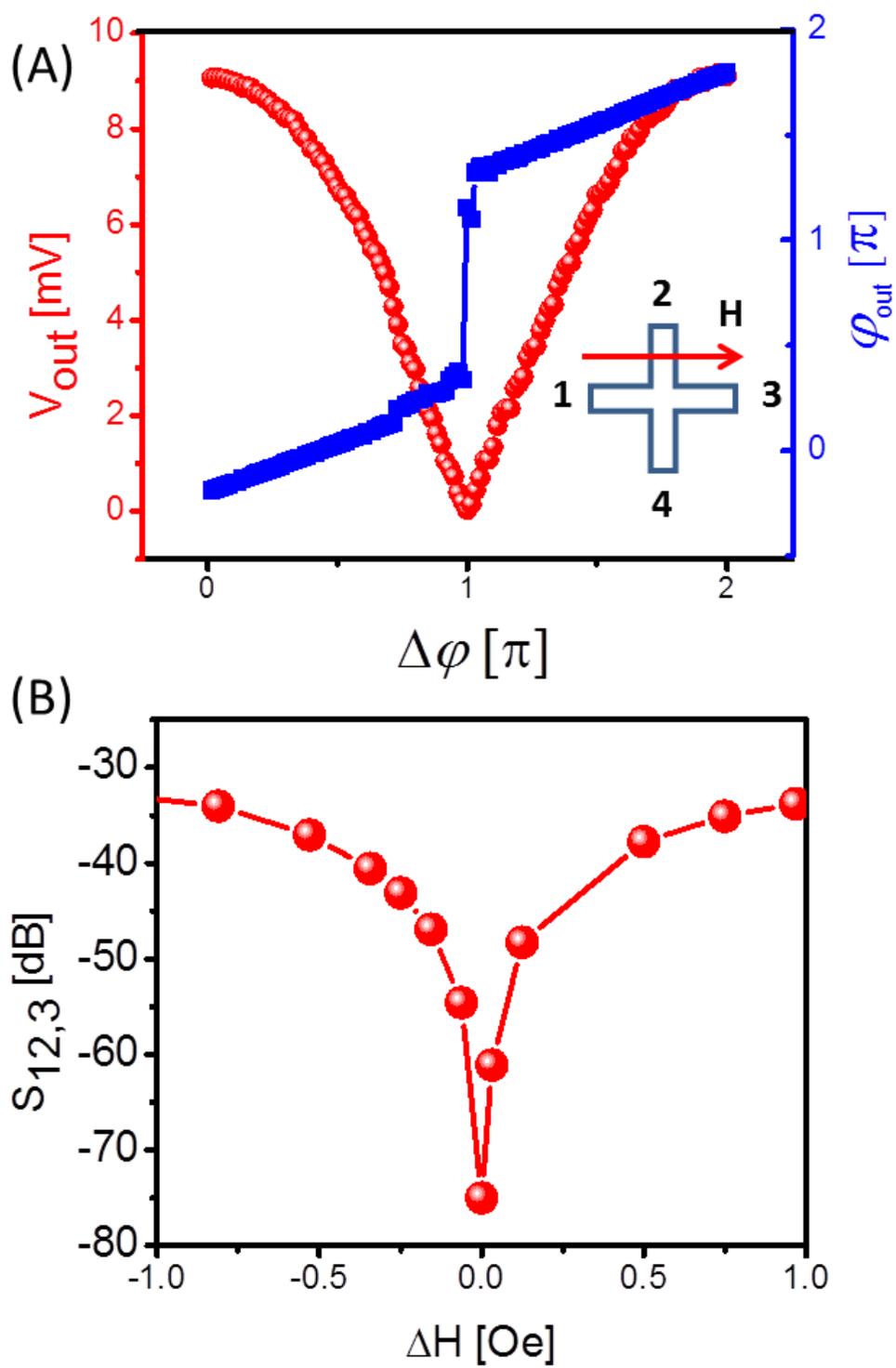

Figure 4

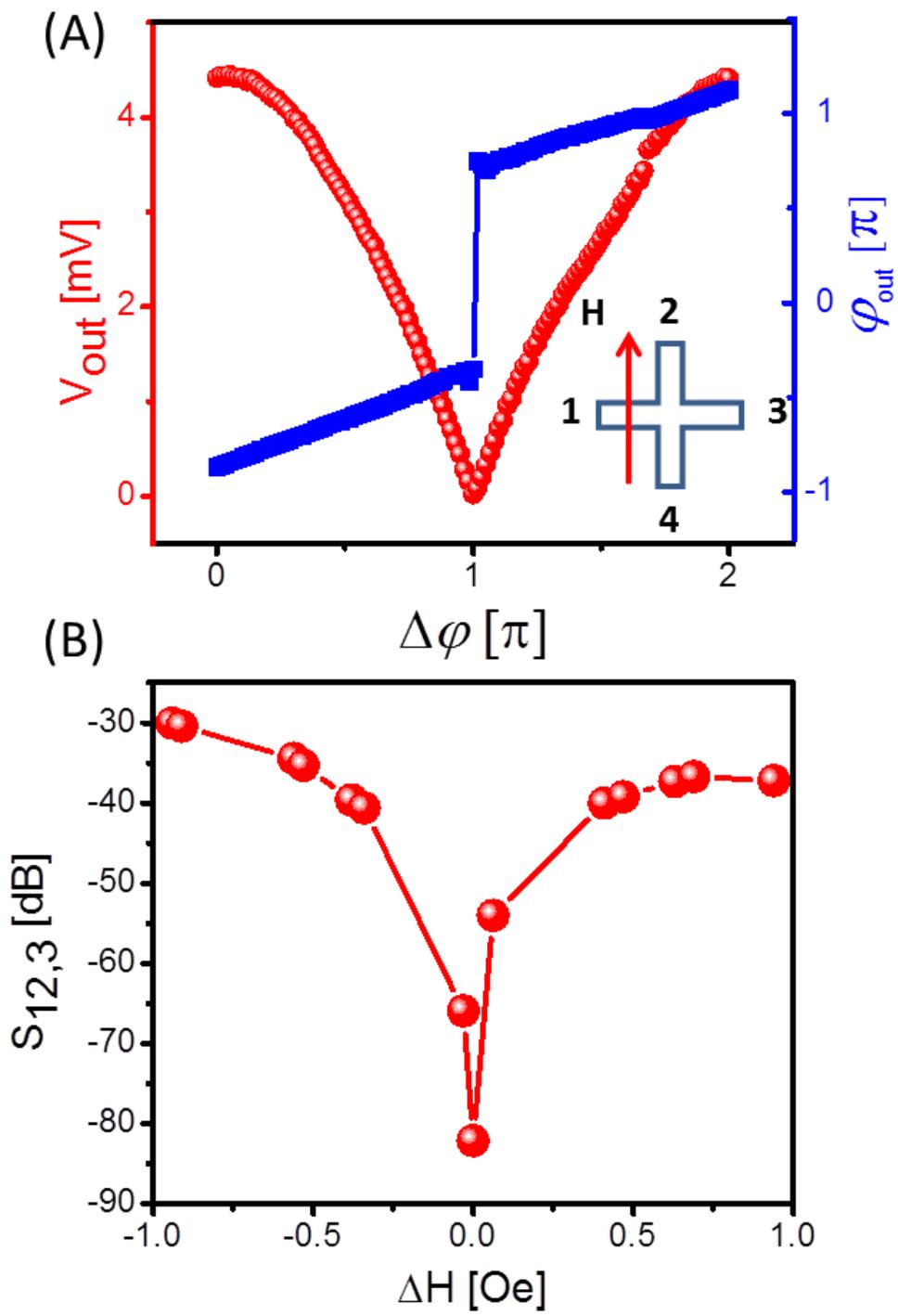

Figure 5

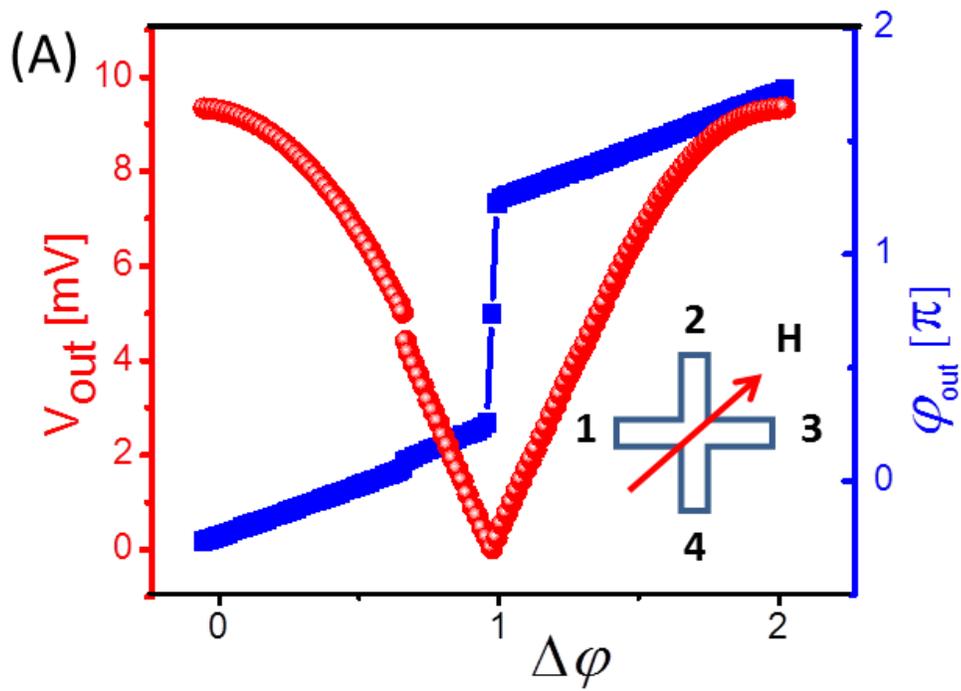

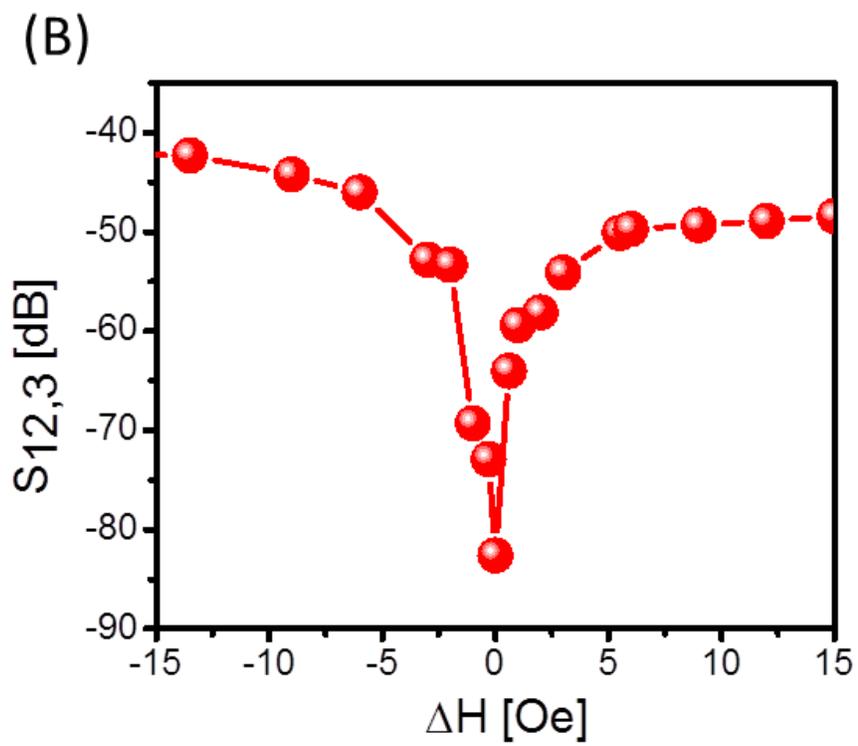

Figure 6

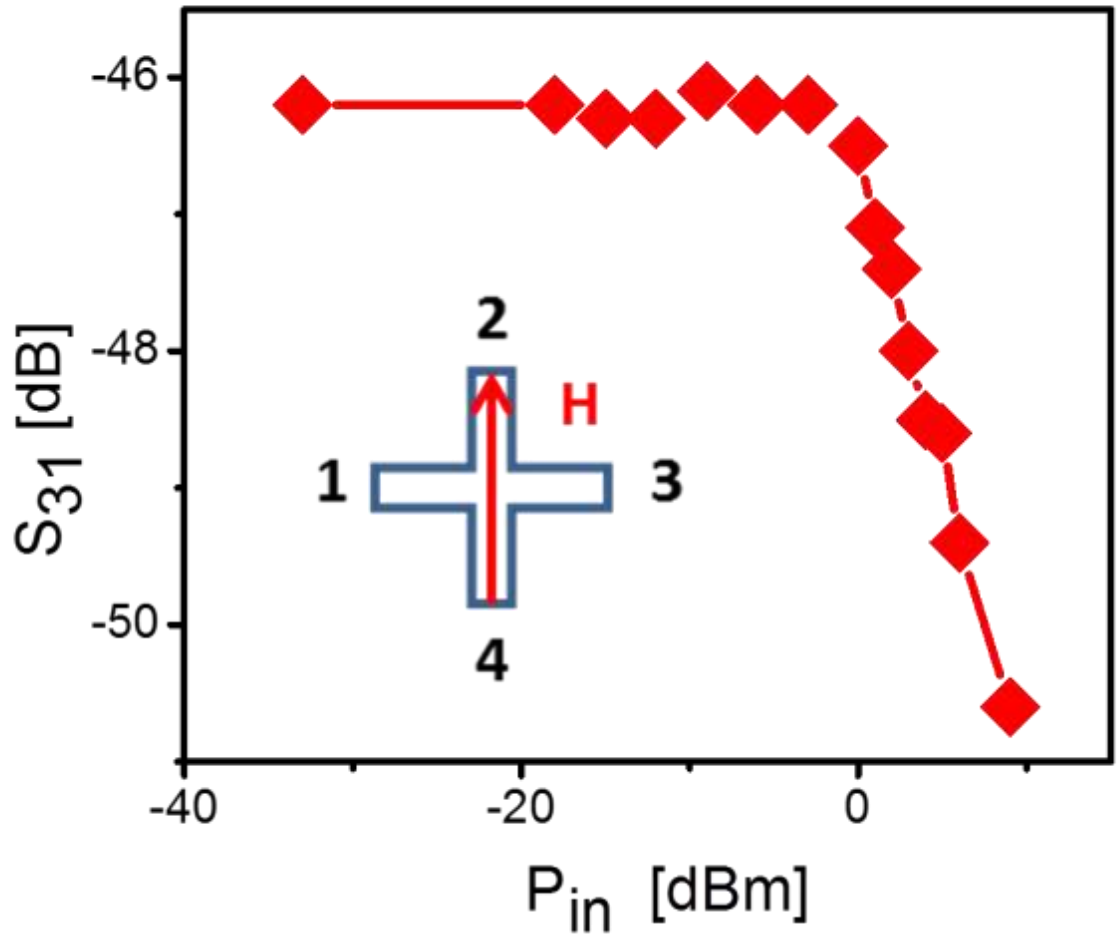

Figure 7